\documentclass[aps,prd]{revtex4}
\usepackage{epsfig,epsf}
\usepackage{amsmath}
\usepackage{amsthm}
\usepackage{amsfonts}
\usepackage{amssymb}
\usepackage{dsfont}
\usepackage{multirow}
\usepackage{appendix}
\usepackage{slashed}
\usepackage[active]{srcltx}
\usepackage{psfrag}
\usepackage{subfigure}
\usepackage[colorlinks,citecolor=blue,urlcolor=red,linkcolor=red]{hyperref}
\usepackage[colorlinks,citecolor=blue,urlcolor=red,linkcolor=red]{hyperref}
\begin{document}

\title{\Large{\bf { Analysis of $D^*_sD^*K^*$ and $ D_{s1} D_1 K^*$ vertices  in three-point sum rules}}}

\author{\small
    \vskip5mm M. Janbazi$^1$\footnote {e-mail: mehdijanbazi@yahoo.com },
    R. Khosravi$^2$
    and E. Noori$^{3}$ }
\affiliation{\emph{$^1$Young Researchers and Elites Club, Shiraz Branch, Islamic Azad University, Shiraz, Iran \\
$^2$Department of Physics, Isfahan University of
Technology, Isfahan 84156-83111, Iran \\
$^3$Young Researchers and Elite Club, Karaj Branch, Islamic Azad University, Karaj, Iran} }

\begin{abstract}
In this study, the coupling constant of $D^*_sD^*K^*$ and $D_{s1}D_1K^*$  vertices were determined
within the three-point Quantum chromodynamics sum rules method with and without consideration of the
$SU_{f}(3)$ symmetry. The coupling constants were calculated for off-shell charm and K$^*$ cases. Considering the non-perturbative effect of the correlation function, as the most important contribution, the quark-quark,
quark-gluon, and gluon-gluon condensate corrections were estimated and were compared with other predictive methods.\\
\\
\textit{keywords}: Strong form factors, Coupling constants, Quantum chromodynamics, Mesons
\end{abstract}


\maketitle

\section{Introduction}

Considerable attention has been focused on the strong form factors
and coupling constants of meson vertices in the context of Quantum
chromodynamics (QCD) since the last decade. In high energy physics,
understanding the functional form of the strong form factors plays
very important role in explaining of the meson interactions.
Therefore, accurate determination of the strong form factors and
coupling constants associated with the vertices involving mesons,
has been attracted great interest in recent studies of the high
energy physics.

Quantum chromodynamics sum rules (QCDSR) formalism have been used
extensively to study about the " exotic " mesons made of quark-
gluon hybrid ($ q\bar{q}g $), tetraquark states ($ q\bar{q}q\bar{q}
$), molecular states of two ordinary mesons, glueballs \cite{exotic}
and vertices involving charmed mesons such as $D^* D^*
\rho$\cite{FSNavarra1,MEBracco}, $D^* D \pi$
\cite{FSNavarra1,MNielsen}, $D D \rho$\cite{MChiapparini}, $D^* D
\rho$\cite{Rodrigues3}, $D D J/\psi$ \cite{RDMatheus},  $D^* D
J/\psi$ \cite{RRdaSilva}, $D^*D_sK$, $D^*_sD K$, $D^*_0 D_s K$,
$D^*_{s0} D K$ \cite{SLWang}, $D^*D^* P$, $D^*D V$, $D D V$
\cite{ZGWang}, $D^* D^* \pi$ \cite{FCarvalho}, $D_s D^* K$, $D_s^* D
K$ \cite{ALozea}, $D D \omega$ \cite{LBHolanda}, $D_{s0}DK$ and
$D_0D_sK$  \cite{Wang122}, $D_{s1}D^*K$, $D_{s1}D^*K_0^*$
\cite{Wang3434,Ghahramany},  $D_s D_s V$, $D^{*}_s D^{*}_s V$
\cite{KJ,KJ2}  and $D_1D^*\pi, D_1D_0\pi, D_1D_1\pi$ \cite{Janbazi}.

In this study, 3-point sum rules (3PSR) method is used to calculate
the strong form factors and coupling constants of the $D^*_{s} D^{*}
K^*$ and $D_{s1} D_{1} K^*$ vertices. The 3PSR correlation function
is investigated from the phenomenological and the theoretical points
of view. Regarding the phenomenological (physical) approach, the
representation can be expressed in terms of hadronic degrees of
freedom which can be considered as responsible of the introduction
of the form factors, decay constant and masses. The theoretical
(QCD) approach usually can be divided into two main contributions as
perturbative and non-perturbative. In this approach, the quark-gluon
language and Wilson operator product expansion (OPE) are usually
used to evaluate the correlation function in terms of the QCD
degrees of freedom such as quark condensate, gluon condensate,
\textit{etc}. Equating the two sides and applying the double Borel
transformations with respect to the momentum of the initial and
final states to suppress the contribution of the higher states, and
continuum, the strong form factors can be estimated.

The effective Lagrangian of the interaction for the $D^*_s D^* K^*$
and $D_{s1} D_{1} K^*$ vertices can be written as \cite{BraChia}
\begin{eqnarray}\label{eq11}
{\cal L}_{D_s^* D^* K^*}&=& ig_{D_s^* D^* K^*}[{D_s^*}^\mu(\partial_{\mu}{K^*}^\nu \bar D^*_{\nu}-{K^*}^\nu\partial_{\mu}\bar D^*_{\nu})+(\partial_{\mu}{D^*_s}^\nu K^{*}_{\nu}
-{D_s^*}^\nu\partial_{\mu}K^{*}_{\nu}){\bar D}^{*\mu}\nonumber\\&&+ {K^*}^\mu
({D_s^*}^{\nu}\partial_{\mu} \bar D^*_\nu  -
\partial_{\mu}{D^*_s}^\nu  \bar D^*_\nu) ],
\end{eqnarray}
\begin{eqnarray}\label{eq11}
{\cal L}_{D_{s1} D_{1} K^*}&=& ig_{D_{s1} D_{1} K^*}[{D_{s1}}^\mu(\partial_{\mu}{K^*}^\nu \bar D_{1\nu}-{K^*}^\nu\partial_{\mu}\bar D_{1\nu})+(\partial_{\mu}{D_{s1}}^\nu K^{*}_{\nu}
-{D_{s1}}^\nu\partial_{\mu}K^{*}_{\nu}){\bar D}_{1}^{\mu}\nonumber\\&&+ {K^*}^\mu
({D_{s1}}^{\nu}\partial_{\mu} \bar D_{1\nu}-
\partial_{\mu}{D_{s1}}^\nu  \bar D_{1\nu}) ],
\end{eqnarray}
where $g_{D^*_s D^* K^*}$ and $g_{D_{s1} D_{1} K^*}$
is the strong form factor. Using the introduced form of the Lagrangian, the elements
related to the $D^*_s D^* K^*$ and $D_{s1} D_{1} K^*$
vertices can be derived in terms of the strong form factor as
\begin{eqnarray}\label{eq12}
\langle D^*(p,\varepsilon) D_s^*(p', \varepsilon') |  K^*(q,\varepsilon'')
\rangle &=& ig_{D^*_s D^*
    K^*}(q^2)\times [{(q^\alpha+p'^\alpha)} g^{\mu\nu}- {(q^{\mu}+p^{\mu})} g^{\nu\alpha}+{q}^\nu g^{\alpha\mu}]\nonumber\\&\times&\varepsilon_{\alpha
}(p)\varepsilon'_{\mu}(p')\varepsilon''_{\nu}(q),
\end{eqnarray}
\begin{eqnarray}\label{eq1211}
\langle D_1(p,\varepsilon) D_{s1}(p', \varepsilon') |  K^*(q,\varepsilon'')
\rangle &=& ig_{D_{s1} D_{1} K^*}(q^2)\times [{(q^\alpha+p'^\alpha)} g^{\mu\nu}- {(q^{\mu}+p^{\mu})} g^{\nu\alpha}+{q}^\nu g^{\alpha\mu}]\nonumber\\&\times&\varepsilon_{\alpha
}(p)\varepsilon'_{\mu}(p')\varepsilon''_{\nu}(q),
\end{eqnarray}
where $q=p-p'$.

The organization of the paper is as follows: In Section II, the
quark-quark, quark-gluon and gluon-gluon condensate contributions,
considering the non-perturbative effects of the Borel transform
scheme, are discussed in order to calculate the strong form factors
of the $D^*_{s} D^{*} K^*$ and $D_{s1} D_{1} K^*$  vertices in the
framework of the 3PSR. The numerical analysis of the strong form
factors estimation as well as the coupling constants, with and
without consideration of the $SU_{f}(3)$ symmetry is described in
Section III and the conclusion is made in section IV.

\section{The STRONG FORM FACTOR OF $D^*_{s}    D^{*} K^*$   and $D_{s1}    D_1 K^*$ vertices}

\begin{figure}[!th]
\centering
\includegraphics[scale=0.5]{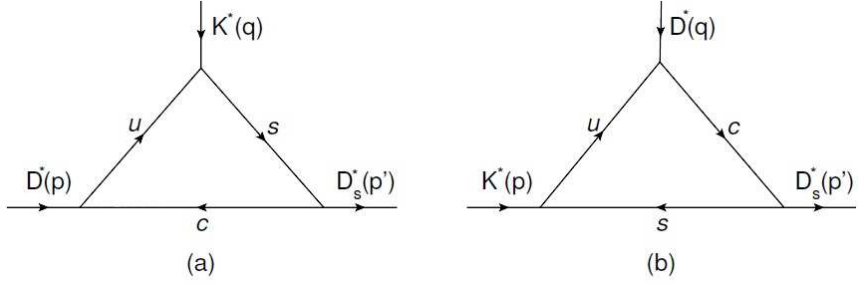}
\caption{Perturbative diagrams for off-shell $K^*$
        (a) and off-shell $D^*$   (b).}
\label{F1}
\end{figure}

To compute the strong form factor of the $D^*_sD^*K^*$  and
$D_{s1}D_1K^*$ vertices via the 3PSR, we start with the correlation
function. When the $K^*$ meson is off-shell, the correlation
function can be written in the following form
\begin{eqnarray}\label{eq21}
\Pi^{K^*}_{\mu\nu\alpha}(p, p')&=& i^2 \int d^4x  d^4y e^{i(p'x-py)}\langle 0
|\mathcal{T}\left\{j_{\mu}^{D^*_s}(x) {j_{\nu}^{K^*}}^{\dagger}(0)
{j_{\alpha}^{D^*}}^{\dagger}(y)\right\}| 0 \rangle,
\end{eqnarray}
\begin{eqnarray}\label{eq2111}
\Pi^{K^*}_{\mu\nu\alpha}(p, p')&=& i^2 \int d^4x  d^4y
e^{i(p'x-py)}\langle 0 |\mathcal{T}\left\{j_{\mu}^{D_{s1}}(x)
{j_{\nu}^{K^*}}^{\dagger}(0)
{j_{\alpha}^{D_1}}^{\dagger}(y)\right\}| 0 \rangle.
\end{eqnarray}
For off-shell charm meson, the correlation function can be written as
\begin{eqnarray}\label{eq22}
\Pi^{D^*}_{\mu\alpha\nu}(p, p') &=& i^2 \int d^4x  d^4y e^{i(p'x-py)}\langle 0
|\mathcal{T}\left\{ j_{\mu}^{D^*_s}(x){j_{\alpha}^{D^*}}^{\dagger}(0){j_{\nu}^{K^*}}^{\dagger}(y)
\right\}| 0 \rangle,
\end{eqnarray}
\begin{eqnarray}\label{eq2211}
\Pi^{D_1}_{\mu\alpha\nu}(p, p') &=& i^2 \int d^4x  d^4y e^{i(p'x-py)}\langle 0
|\mathcal{T}\left\{ j_{\mu}^{D_{s1}}(x){j_{\alpha}^{D_1}}^{\dagger}(0){j_{\nu}^{K^*}}^{\dagger}(y)
\right\}| 0 \rangle,
\end{eqnarray}
where $j_{\mu}^{D^*_{s}}=\bar c \gamma_{\mu} s,~j_{\alpha}^{D^*}=\bar c \gamma_{\alpha} u$, $j_{\mu}^{D_{s1}}=\bar c \gamma_{\mu}\gamma_5 s,~j_{\alpha}^{D_1}=\bar c \gamma_{\alpha} \gamma_5 u$ and
$j^{K^*}_{\nu}=\bar u \gamma_{\nu} s $ are interpolating currents
with the same quantum numbers of $D^*_{s},~D^*$, $D_{s1},~D_1$, and $K^*$ mesons. As described in Fig. \ref{F1}, $\mathcal{T}$, $p$ and $p'$ are time ordering product and four momentum of the initial and final mesons respectively.

Considering the OPE scheme in the phenomenological approach, the
correlation functions (Eqs. (\ref{eq21})- (\ref{eq2211})) can be
written in terms of several tensor structures which their
coefficients are found using the sum rules.
 It is clear from Eqs. (\ref{eq12} \& \ref{eq1211}) that the  form factor $g_{D^*_s D^* K^*}$ is used for the
fourth Lorentz structure which can be extracted from the sum rules.
We choose the Lorentz structure because of its fewer ambiguities in
the 3PSR approach, i.e. less influence of higher dimension of the
condensates and better stability as function of the Borel mass
parameter \cite{FSNavarra1}. For these reasons, the
$g_{\mu\alpha}q_{\nu}$ structure is chosen which is  assumed to
better formulate the problem.

In order to calculate the phenomenological part of the correlation
functions in Eqs. (\ref{eq21})- (\ref{eq2211}), three complete sets
of intermediate states with the same quantum number as the currents
$j_{\mu}^{D_s^*}$, $j_{\alpha}^{D^*}$, $j_{\mu}^{D_{s1}}$,
$j_{\alpha}^{D_1}$ and  $j_{\nu}^{K^*}$ are selected. The matrix
elements $\langle 0 | j_{\mu}^{D_s^*} | D^*_s(p,\varepsilon)
\rangle$,  $\langle 0 | j_{\alpha}^{D^*} | D^*(p) \rangle$, $\langle
0 | j_{\mu}^{D_{s1}} | D_{s1}(p,\varepsilon) \rangle$,  $\langle 0 |
j_{\alpha}^{D_1} | D_1(p) \rangle$,  and $\langle 0 | j_{\nu}^{K^*}
| K^*(q,\varepsilon) \rangle$ are defined as
\begin{eqnarray}\label{eq29}
\langle 0 | j_{\mu}^{V} | V(p,\varepsilon) \rangle &=& m_{V}
f_{V} \varepsilon_{\mu}(p),
\end{eqnarray}
where $m_{V}$ and $f_{V}$  are the masses and decay constants of mesons $V(D_s^*,D^*,D_{s1},D_1,K^*)$ and
$\varepsilon_\mu$ is introduced as the polarization vector
of the vector  meson $V(D_s^*,D^*,D_{s1},D_1,K^*)$.

The phenomenological part of the $g_{\mu\alpha}q_{\nu}$ structure
associated with the $D^*_sD^*K^*$ vertex for off-shell $D^*$ and
$K^*$ mesons can be expressed as
\begin{eqnarray}\label{eq210}
\Pi_{\mu\nu\alpha}^{D^*} &=& -g^{D^*}_{D^{*}_{s} D^* K^* }(q^2)\frac{m_{K^*}
    f_{K^*}f_{D^*}f_{D_s^*} (3m_{D^*}^2+m_{K^*}^2-q^2)}{ 2 m_{D_s^*}(q^2-m_{D^*}^{2})(p^2-m_{K^*}^2)(p'^2-m_{D^*_s}^2)}+\mbox{...}\, ,\nonumber\\
\Pi_{\mu\nu\alpha}^{K^*}  &=&g^{K^*}_{D^{*}_{s} D^* K^* }(q^2)\frac{m_{K^*} f_{K^*}f_{D^*}f_{D_s^*} (3m_{D_s^*}^2+m_{D^*}^2-q^2)}{2 m_{D_s^*} (q^2-m_{K^*}^{2})(p^2-m_{D^*}^2)(p'^2-m_{D^*_s}^2)
    }+\mbox{...}~.
\end{eqnarray}

The phenomenological part of the $g_{\mu\alpha}q_{\nu}$ structure
associated with the $D_{s1}D_1K^*$ vertex for off-shell $D_1$ and
$K^*$ mesons can be expressed as
\begin{eqnarray}\label{eq21011}
\Pi_{\mu\nu\alpha}^{D_1} &=& -g^{D_1}_{D_{s1} D_1 K^* }(q^2)\frac{m_{K^*}
    f_{K^*}f_{D_1}f_{D_{s1}} (3m_{D_1}^2+m_{K^*}^2-q^2)}{ 2 m_{D_{s1}}(q^2-m_{D_1}^{2})(p^2-m_{K^*}^2)(p'^2-m_{D_{s1}}^2)}+\mbox{...}\, ,\nonumber\\
\Pi_{\mu\nu\alpha}^{K^*}  &=&g^{K^*}_{D_{s1} D_1 K^*
}(q^2)\frac{m_{K^*} f_{K^*}f_{D_1}f_{D_{s1}}
(3m_{D_{s1}}^2+m_{D_1}^2-q^2)}{2 m_{D_{s1}}
(q^2-m_{K^*}^{2})(p^2-m_{D^*}^2)(p'^2-m_{D_{s1}}^2) }+\mbox{...}~.
\end{eqnarray}

Using the operator product expansion in Euclidean
region and assuming $p^2,p'^2\to -\infty$, one can calculate the QCD side of
the correlation function (Eqs. (\ref{eq21})- (\ref{eq2211}))
which contains perturbative and non-perturbative terms.
Using  the double dispersion relation  for the coefficient of the
Lorentz structure  $g_{\mu\alpha}q_{\nu}$  appearing in the correlation function (Eqs. (\ref{eq12}) \& (\ref{eq1211})), we get
\begin{eqnarray}\label{eq28}
\Pi_{per}^{M} (p^2, p'^2, q^2)=
-\frac{1}{4 \pi^2} \int
ds\int ds'\frac{\rho^{M}(s, s',
    q^2)}{(s-p^2)(s'-p'^2)}+\mbox{subtraction  terms},
\end{eqnarray}
where $\rho^{M}(s, s', q^2)$ is spectral density and  $M$ stands for
off-shell charm and $K^*$ mesons. The Cutkosky’s rule allows us to
obtain the spectral densities of the correlation function for the
Lorentz structure appearing in the correlation function. As shown in
Fig. \ref{F1}, the leading contribution comes from the perturbative
term. As a result, the spectral densities are obtained in the case
of the double discontinuity in Eq. (\ref{eq28}) for the vertices;
see Appendix-A.

\begin{figure}[!hbp]
\centering
\includegraphics[scale=0.95]{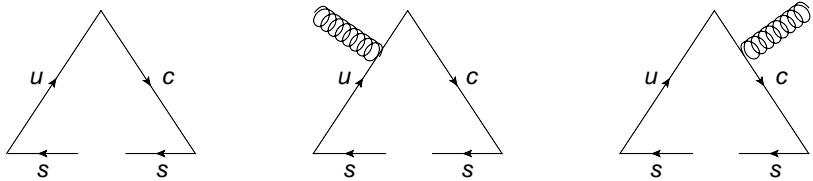}
\caption{Non-perturbative diagrams for the off-shell
        $D^*$ meson.}
\label{F2}
\end{figure}

In order to consider the non-perturbative part of the correlation
functions for the case of spectator light quark (for off-shell charm
meson ), we proceed to calculate the non-perturbative contributions
in the QCD approach which contain the quark-quark and quark-gluon
condensates \cite{Colangelo}. Fig. \ref{F2} describes the important
quark-quark and quark-gluon condensates from the non-perturbative
contribution of the off-shell charm mesons \cite{Colangelo}.

\begin{figure}[!hbp]
\centering
\includegraphics[scale=0.95]{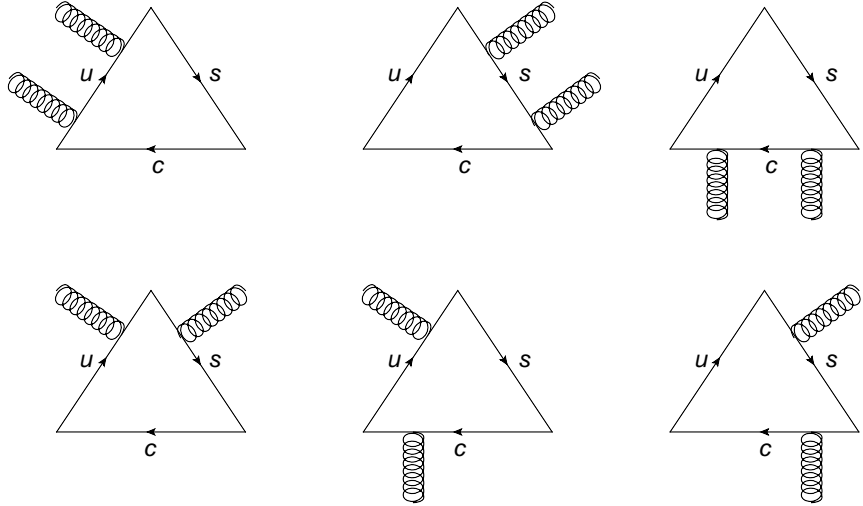}
\caption{Non-perturbative diagrams for the off-shell
        $K^*$ meson.}
\label{F4}
\end{figure}

In the 3PSR frame work, when the  heavy quark  is a spectator (for
off-shell $K^*$ meson), the gluon-gluon contribution can be
considered. Fig. \ref{F4} shows related diagrams of the gluon-gluon
condensate. More details about the non-perturbative contribution
$C^{D^*}_{ D^*_{s} D^*K^*}$ and $C^{D_1}_{ D_{s1} D_1K^*}$(sum
contributions of quark-quark and quark-gluon condensates) and
$C^{K^*}_{ D^*_{s} D^*K^*}$ and $C^{K^*}_{ D_{s1} D_1K^*}$(for
gluon-gluon condensates) corresponding to Figs. \ref{F2} and
\ref{F4} are given in Appendix-B, respectively.

Considering the perturbative and nonperturbative parts of the
correlation function in order to suppress the contributions of the
higher states, the strong form factors can be calculated in the
phenomenological side by equating the two representations of the
correlation function and applying the Borel transformations with
respect to the $p^2(p^2\rightarrow M^2_1)$ and $p'^2(p'^2\rightarrow
M^2_2)$. The equations for the strong form factors
$g^{K^*}_{D^*_sD^*K^* }$  and $g^{D^*}_{D^*_sD^*K^*}$ are obtained
as
\begin{eqnarray}\label{eq212}
g^{K^*}_{D^*_sD^*K^* }(q^2)&=&\frac{2 m_{D_s^*} (q^2-m_{K^*}^{2})}{
    m_{K^*} f_{K^*}f_{D^*}f_{D_s^*} (3m_{D_s^*}^2+m_{D^*}^2-q^2)}~ e^{\frac{m_{D^*}^2}{M_1^2}}e^{\frac{m_{D^*_s}^2}{M_2^2}}
\left\{-\frac{1}{4\pi^2}\int^{s_0^{D^*_s}}_{(m_c+m_s)^2}ds'\right.\nonumber \\
&&\left.
\int^{s_0^{D^*}}_{(m_u+m_c)^2} ds \rho^{K^*}_{D^*_sD^*K^*}(s,s',q^2)
e^{-\frac{s}{M_1^2}}
e^{-\frac{s'}{M_2^2}} - iM^{2}_{1}M^{2}_{2} \left \langle \frac{\alpha_s}{\pi} G^2
\right\rangle \times C^{K^*}_{ D^*_{s} D^*K^*} \right\},\nonumber\\
g^{D^*}_{D^*_sD^*K^*}(q^2)&=&  \frac{2 m_{D_s^*}(q^2-m_{D^*}^{2})}{ m_{K^*}
    f_{K^*}f_{D^*}f_{D_s^*} (3m_{D^*}^2+m_{K^*}^2-q^2)}~e^{\frac{m_{K^*}^2}{M_1^2}} e^{\frac{m_{D^*_s}^2}{M_2^2}}
\left\{-\frac{1}{4\pi^2}\int^{s^{D^*_s}_0}_{(m_c +m_s)^2}ds'\right.\nonumber \\
&&\left.
\int^{s^{K^*}_0}_{(m_u+m_s)^2} ds \rho^{D^*}_{D^*_s D^*K^*}(s,s',q^2)
e^{-\frac{s}{M_1^2}} e^{-\frac{s'}{M_2^2}} +M_1^2 M_2^2~  \langle s\bar s\rangle
\times C^{D^*}_{ D^*_{s} D^*K^*}) \right\}.
\end{eqnarray}

The equations describing the strong form factors
$g^{K^*}_{D_{s1}D_1K^* }$ and $g^{D_1}_{D_{s1}D_1K^*}$ can be
written as
\begin{eqnarray}\label{eq213}
g^{K^*}_{D_{s1}D_1K^* }(q^2)&=&\frac{2 m_{D_{s1}} (q^2-m_{K^*}^{2})}{
    m_{K^*} f_{K^*}f_{D_1}f_{D_{s1}} (3m_{D_{s1}}^2+m_{D_1}^2-q^2)}~ e^{\frac{m_{D_1}^2}{M_1^2}}e^{\frac{m_{D_{s1}}^2}{M_2^2}}
\left\{-\frac{1}{4\pi^2}\int^{s_0^{D_{s1}}}_{(m_c+m_s)^2}ds'\right.\nonumber \\
&&\left.
\int^{s_0^{D_1}}_{(m_u+m_c)^2} ds \rho^{K^*}_{D_{s1}D_1K^*}(s,s',q^2)
e^{-\frac{s}{M_1^2}}
e^{-\frac{s'}{M_2^2}} - iM^{2}_{1}M^{2}_{2} \left \langle \frac{\alpha_s}{\pi} G^2
\right\rangle \times C^{K^*}_{ D_{s1}D_1K^*} \right\},\nonumber\\
g^{D_1}_{D_{s1}D_1K^*}(q^2)&=&  \frac{2 m_{D_{s1}}(q^2-m_{D_1}^{2})}{ m_{K^*}
    f_{K^*}f_{D_1}f_{D_{s1}} (3m_{D_1}^2+m_{K^*}^2-q^2)}~e^{\frac{m_{K^*}^2}{M_1^2}} e^{\frac{m_{D_{s1}}^2}{M_2^2}}
\left\{-\frac{1}{4\pi^2}\int^{s^{D_{s1}}_0}_{(m_c +m_s)^2}ds'\right.\nonumber \\
&&\left. \int^{s^{K^*}_0}_{(m_u+m_s)^2} ds \rho^{D_1}_{D_{s1}
D_1K^*}(s,s',q^2) e^{-\frac{s}{M_1^2}} e^{-\frac{s'}{M_2^2}} +M_1^2
M_2^2~  \langle s\bar s\rangle \times C^{D_1}_{ D_{s1}D_1K^*}
\right\},
\end{eqnarray}
where the quantities $s^{D^*_s}_0$,  $s^{D^*}_0$ $s^{D_{s1}}_0$,
$s^{D_1}_0$ and $s^{K^*}_0$ are introduced as the continuum
thresholds in $D^*_s$, $D^*$ $D_{s1}$, $D_1$ and $K^*$ mesons
respectively and $\rho^{K^*}_{D^*_{s} D^*K^*}$,  $\rho^{D^*}_{
D^*_{s} D^*K^*}$, $C^{K^*}_{ D^*_{s} D^*K^*}$, $C^{D^*}_{ D^*_{s}
D^*K^*}$, $\rho^{K^*}_{D_{s1} D_1K^*}$,  $\rho^{D_1}_{
D_{s1}D_1K^*}$, $C^{K^*}_{ D_{s1}D_1K^*}$, and $C^{D_1}_{
D_{s1}D_1K^*}$ are defined in Appendix-A and Appendix-B.

\section{NUMERICAL ANALYSIS}

In order to numerically estimate the strong form factors and
coupling constants of the vertices $D^*_{s} D^{*} K^*$ and $D_{s1}
D_{1} K^*$, the values of the quark and meson masses are chosen as:
$m_s = 0.14\pm0.01~\rm GeV$, $m_{K^*}=0.89~ \rm GeV$,
$m_{D^{*}_{s}}=2.11~\rm GeV$, $m_{D_{s1}}=2.46~\rm GeV$, and
$m_{D_{1}}=2.42~\rm GeV$ \cite{PDG2012}. Moreover,  the leptonic
decay constants of the vertices are:  $f_{K^*}=220\pm5$
\cite{PDG2012}, $f_{D^{*}_{s}}=314\pm 19$ \cite{Gelhausen},
$f_{D^{*}}=242\pm 12$  \cite{Gelhausen},  $f_{D_{s1}}=225\pm 20$
\cite{Thoma} and $f_{D_{1}}=219\pm 11$ \cite{Bazavov}.

There are four auxiliary parameters containing the Borel mass
parameters $M_1$ and $M_2$ and continuum thresholds $s^{K^*}_{0}$,
$s_{0}^{D^*(D_1)}$ and $s^{D^*_s(D_{s1})}_{0}$ in Eqs. (\ref{eq212}
\& \ref{eq213}). The strong form factors and coupling constants are
physical quantities which are independent of the mass parameters and
continuum thresholds. However, the continuum thresholds are not
completely arbitrary and can be related to the energy of the first
exited state. The values of the continuum thresholds are taken to be
$s^{K^*}_{0}=(m_{K^*}+\delta)^2$,
$s_{0}^{D^*(D_1)}=(m_{D^*(D_1)}+\delta')^2$ and
$s_{0}^{D^*_s(D_{s1})}=(m_{D^*_s(D_{s1})}+\delta')^2$ where $0.50
~\rm GeV\leq \delta \leq 0.90~\rm \rm GeV$ and $0.30 ~\rm GeV\leq
\delta' \leq0.70~\rm \rm GeV$ \cite{FSNavarra1,MNielsen,MEBracco}.

\begin{figure}[!hbp]
\centering
\includegraphics[width=8cm,height=7cm]{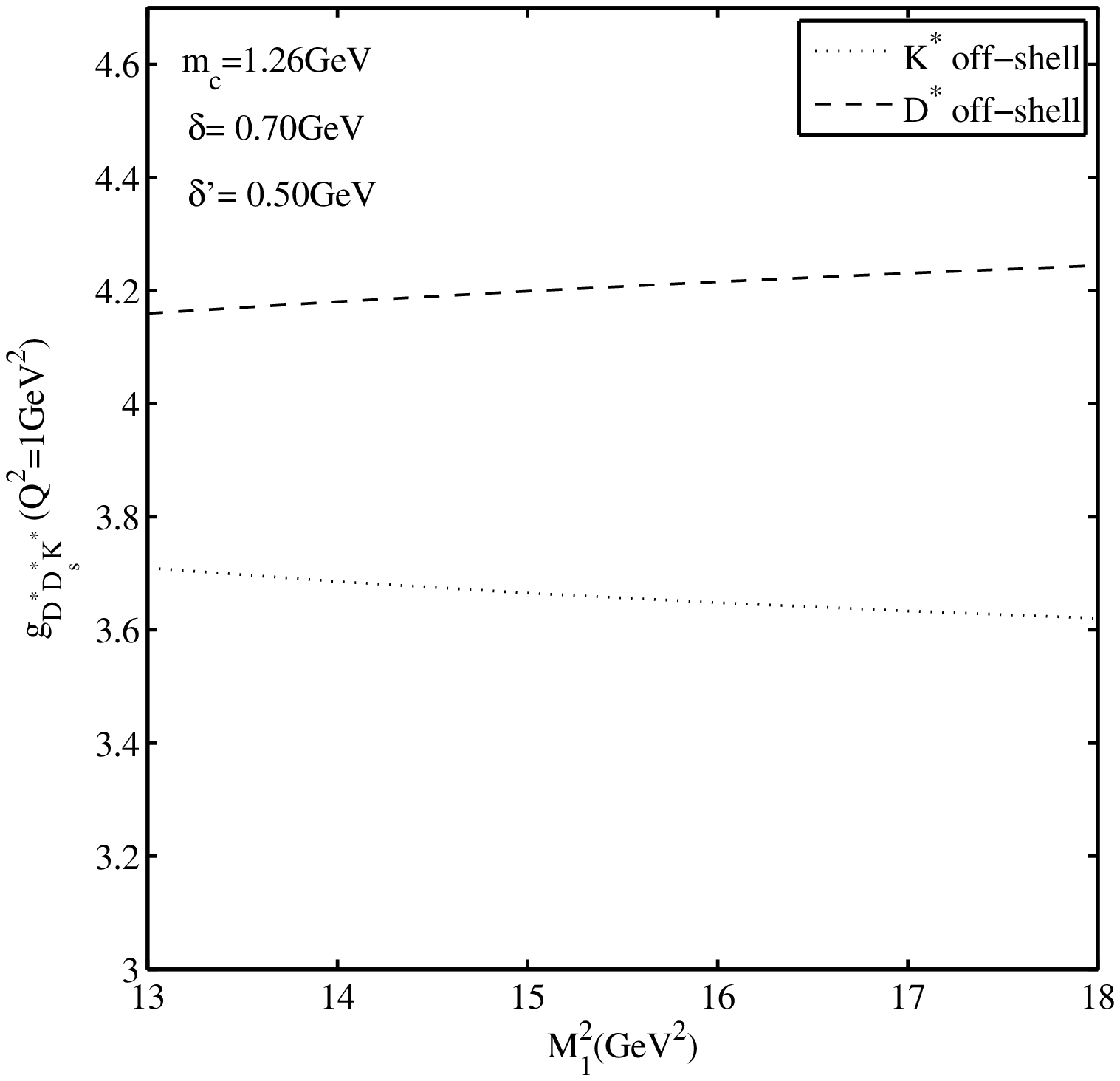}\includegraphics[width=8cm,height=7cm]{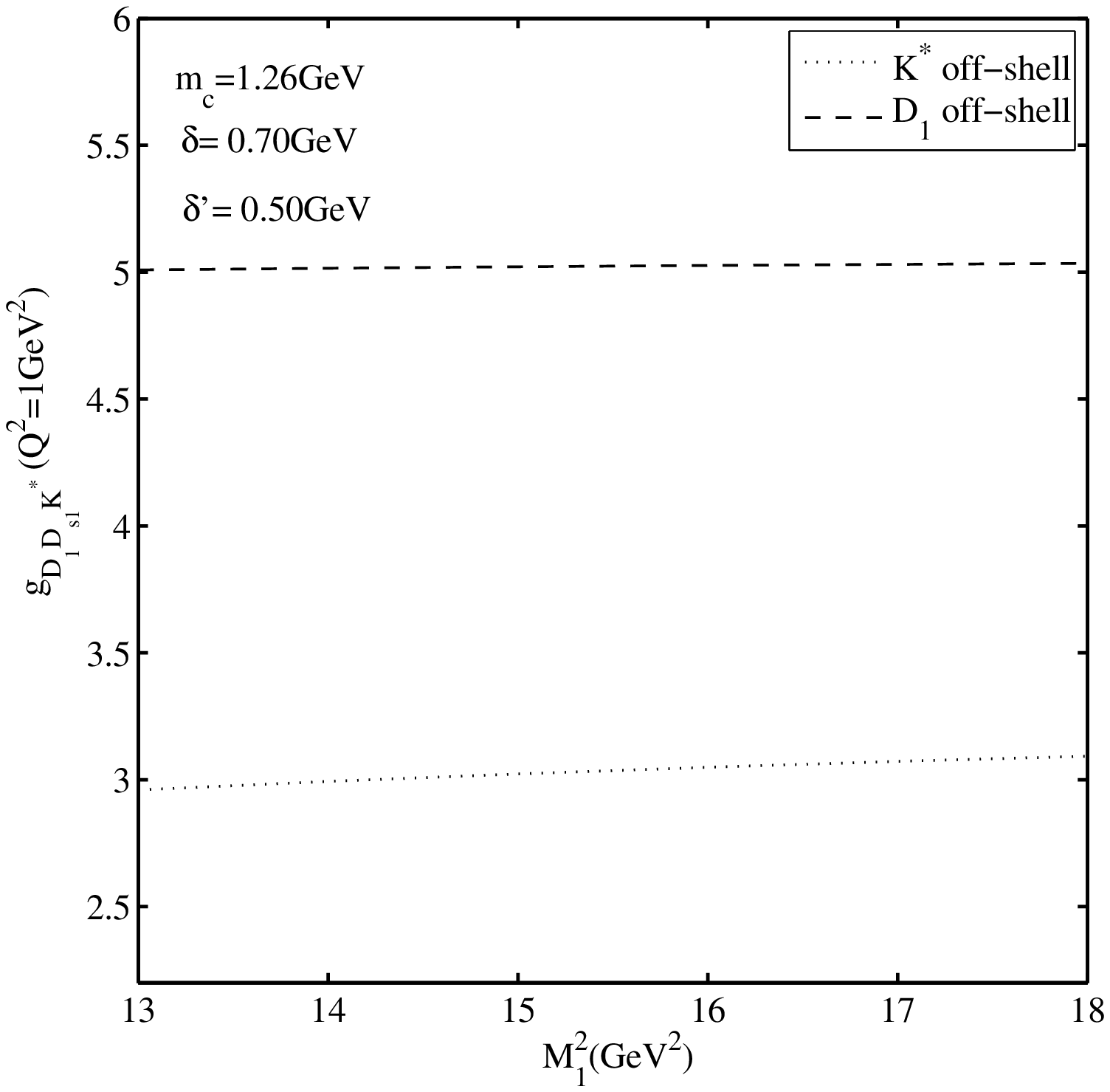}
\caption{The strong form factors $g_{D^*_s D^*
        K^*}$ (left) and $g_{D_{s1} D_1
        K^*}$ (right) as functions of the Borel mass parameter $M_1^2$ for off-shell charm and K$^*$ mesons.}
\label{F301}
\end{figure}

Our results should be almost insensitive to the intervals of the
Borel  parameters. In this work, the Borel masses are related as
$\dfrac{M_1^2}{M_2^2}=\dfrac{m_{K^*}^2+m_{c}^2}{m_{D^*_s(D_{s1})}^2}$
and $M_1^2=M_2^2$ for off-shell charm mesons and K$^*$ respectively
\cite{MChiapparini, Rodrigues3}. The form factors for the
$D^*_sD^*K^*$ and $D_{s1}D_1K^*$ vertices with respect to the Borel
parameters $M_1^2$ are shown in Fig. \ref{F301}. It is found from
the figure that the stability of the form factors, as function of
Borel parameters, is good in the region of $13 ~{\rm GeV^2} < M_1^2
< 18 ~{\rm GeV^2}$ for off-shell $K^*$ and charm  mesons.

We get $M_1^2=15~\rm GeV^2$, and calculate the strong form factors
$g_{D^*_s D^* K^*}$ in some points of $Q^2$ via the 3PSR formalism.

To extract the coupling constants from the form factors, it is
needed to extend the $Q^2$ dependency of the strong form factors to
the ranges that the sum rule results are not valid. Therefore, we
fitted two sets of points (boxes and circles ) imposing the
condition that the two resulting parameterizations lead to the same
result for $Q^2 = -m_m^2$, where $m_m$ is the mass of the off-shell
mesons. This procedure is sufficient to reduce the uncertainties. It
is found that the sum rule predictions of the form factors in Eqs .
(\ref{eq212}  \& \ref{eq213}) are well fitted to the function
\begin{eqnarray*}\label{eq33}
g(Q^2)=A~e^{-Q^2/B}.
\end{eqnarray*}
The values of the parameters $A$ and $B$ are given in Table
\ref{T32}.
\begin{table}
    \caption{Parameters appearing in the fit functions for the
          $D^*_sD^*K^*$ and $D_{s1}D_1K^*$ vertices for various $m_c$ and
        $(\delta, \delta')$, where  $(\delta_1,\delta'_1)=(0.50,0.30),
        ~(\delta_2,\delta'_2)=(0.70,0.50)$ and
        $(\delta_3,\delta'_3)=(0.90,0.70) ~\rm GeV$.}\label{T32}
    \begin{ruledtabular}
        \begin{tabular}{ccccccccc}
            &$\mbox{set I}$&$$$$&$$&$$&$$&$$&$\mbox{set II}$&\\
            \hline
            $\mbox{Form factor}$&$A(\delta_1,\delta'_1)$&$B(\delta_1,\delta'_1)$&$A(\delta_2,\delta'_2)$&$B(\delta_2,\delta'_2)$&$A(\delta_3,\delta'_3)$&$B(\delta_3,\delta'_3)$&$A(\delta_2,\delta'_2)$&$B(\delta_2,\delta'_2)$\\
            $g^{K^*}_{D^*_sD^*K^*}(Q^2)$&4.04&183.10&4.95&197.49&5.93&215.62&4.43&266.52 \\
            $g^{D^*}_{D^*_sD^*K^*}(Q^2)$&3.67&61.46&4.42&52.33&5.56&47.96&4.18&57.95 \\
            $g^{K^*}_{D_{s1}D_1K^*}(Q^2)$&4.19&15.24&4.39&21.18&4.68&29.21&4.22&7.69 \\
            $g^{D_1}_{D_{s1}D_1K^*}(Q^2)$&2.62&13.27&3.13&19.47&4.02&25.32&3.03&14.87 \\
        \end{tabular}
    \end{ruledtabular}
\end{table}

\begin{figure}[!hbp]
    \centering
    \includegraphics[width=8cm,height=7cm]{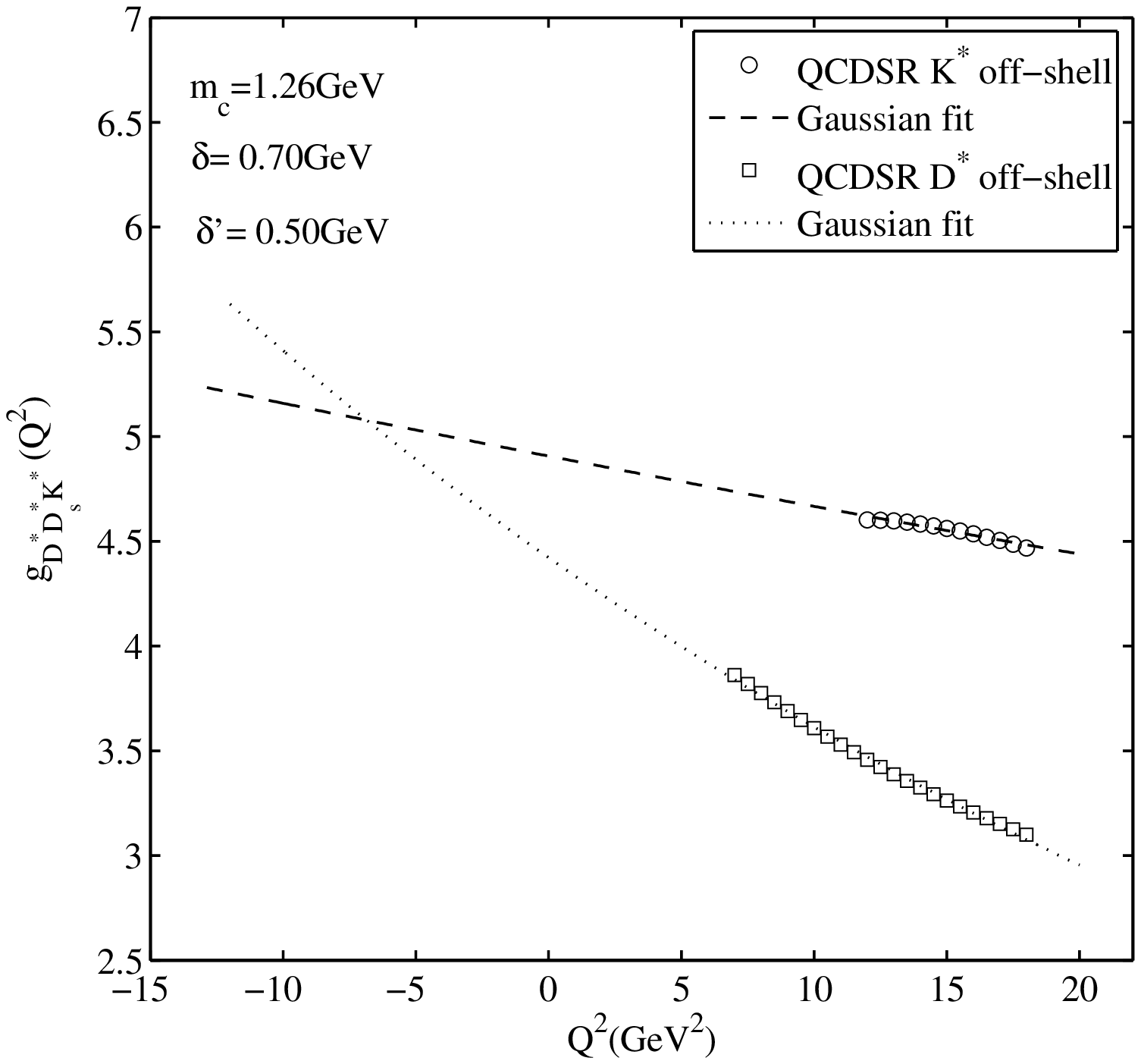}\includegraphics[width=7cm,height=6.5cm]{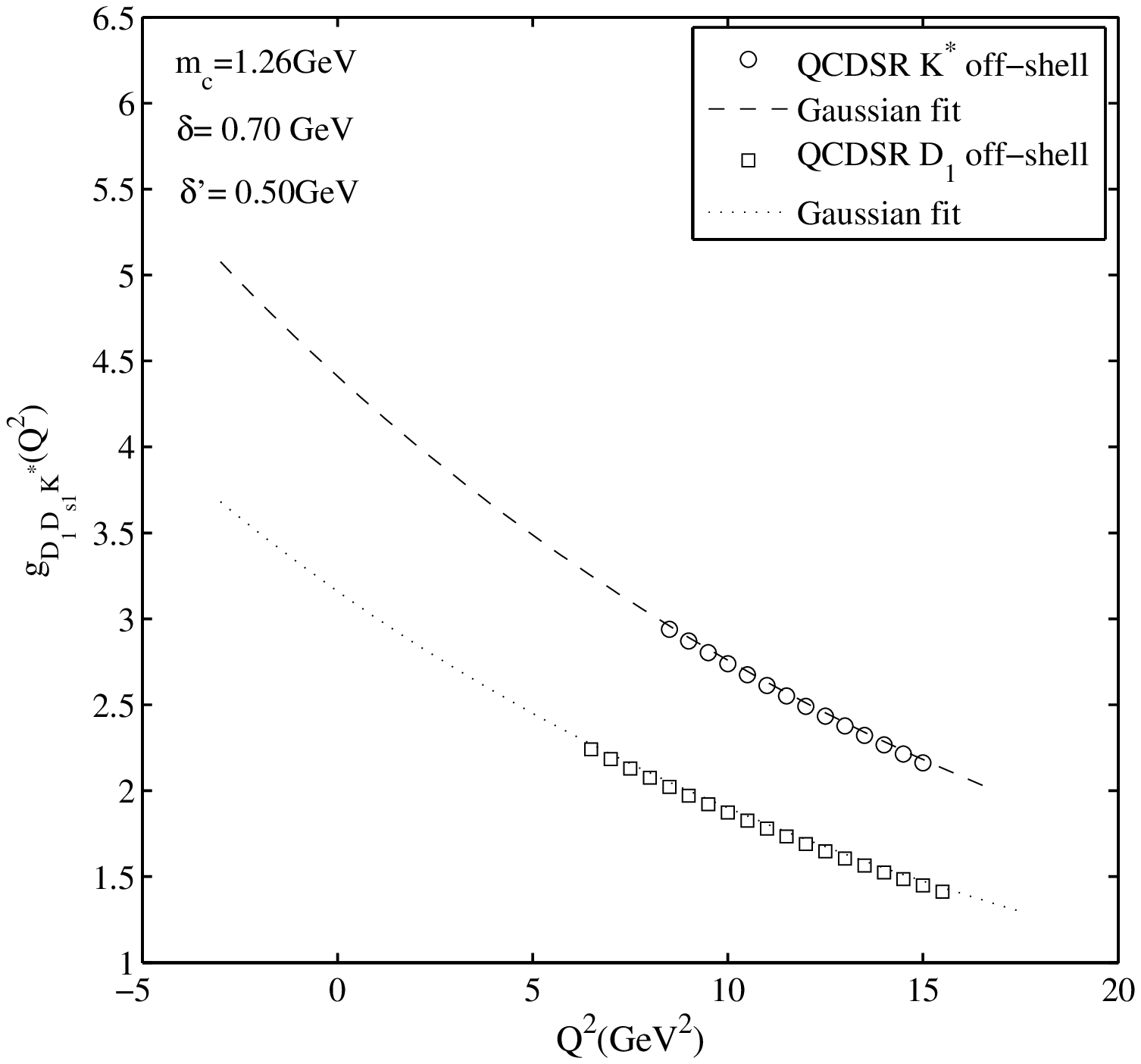}
    \caption{The strong form factors $g_{D^*_s D^*
            K^*}$ (left) and $g_{D_{s1} D_1
            K^*}$ (right) as functions of the $Q^2$ for off-shell charm and K$^*$ mesons.}
    \label{F302}
\end{figure}

Variations of the strong form factors $g_{D^*_s D^* K^*}^{K^*}$ and
$g_{D^*_s D^* K^*}^{D^*}$ for $D^*_s D^* K^*$ vertex and $g_{D_{s1}
D_1 K^*}^{K^*}$ and $g_{{s1} D_1 K^*}^{D_1}$ for $D_{s1} D_1 K^*$
vertex with respect to the $Q^2$ parameter are shown in Fig.
\ref{F302}. The boxes and circles show the results of the numerical
evaluation of the form factors via the 3PSR. It is clear from the
figure that the form factors are in good agreement with the fitted
function.

So-called “harder” is used. In the present analysis we find that
the form factor is harder when the lighter meson is off-shell. This
is in line with the results of our previous work \cite{KJ}, whereas
this is in contrast with other previous calculations quoted by the
authors \cite{Bracco 1111,Rodrigues 111}.

The value of the strong form factors at $Q^2 = -m_m^2$ is defined as
coupling constant. Calculation results of the coupling constant of
the vertices $D^*_{s}D^* K^*$ and $D_{s1}D_1 K^*$ are summarized in
Table \ref{T33}. It should be noted that the coupling constant,
$g_{D^*_sD^*K^*}$ and $g_{D_{s1}D_1K^*}$ are in the unit of $\rm
GeV^{-1}$.
\begin{table}[th]
    \caption{The  coupling constant of the vertices
        ${D^*_sD^*K^*}$ and
        ${D_{s1}D_1K^*}$. }\label{T33}
    \begin{ruledtabular}
        \begin{tabular}{cccccc}
            &$\mbox{set I}$&$$&$\mbox{set II}$&$$\\
            $g$&$\mbox{off-shell charmed }$&$\mbox{off-shell $K^*$ }$&$\mbox{off-shell charmed }$&$\mbox{off-shell $K^*$ }$\\
            \hline
            $g_{D^*_sD^*K^*}$&$4.77\pm0.63$&$4.96\pm0.64$&$4.48\pm0.58$&$4.45\pm0.58$\\
            $g_{D_{s1}D_1K^*}$&$4.22\pm0.55$&$4.56\pm0.59$&$4.48\pm0.58$&$4.67\pm0.62$\\
                \end{tabular}
    \end{ruledtabular}
\end{table}

In order to estimate the error of the calculated parameters,
variations of the Borel parameter, continuum thresholds and leptonic
decay constants, as the most significant reasons of the
uncertainties are considered.

To investigate the value of the strong coupling constant via the
$SU_{f}(3)$ symmetry, the mass of the $s$ quark is ignored in all
equations. Calculated parameters $A$ and $B$ for the
$g_{D^*_sD^*K^*}$ and $g_{D_{s1}D_1K^*}$ vertices, considering
$(\delta,\delta')=(0.70,0.50)~\rm GeV$, are given in Table
\ref{T35}.
\begin{table}[th]
    \caption{Parameters appearing in the fit functions for the
        $g_{D^*_sD^*K^*}$ and $g_{D_{s1}D_1K^*}$ form
        factors in $SU_{f}(3)$ symmetry with $m_c=1.26~\rm GeV$ and
        $(\delta,\delta')=(0.70,0.50)~\rm GeV$.}\label{T35}
    \begin{ruledtabular}
        \begin{tabular}{ccccccc}
            $\mbox{Form factor}$&$A$&$B$&$\mbox{Form factor}$&$A$&$B$\\
            \hline $g^{K^*}_{D^*_sD^*K^*}(Q^2)$&5.01&218.91&$g^{K^*}_{D_{s1}D_1K^*}(Q^2)$&4.56&20.63\\
            $g^{D^*}_{D^*_sD^*K^*}(Q^2)$&4.54&52.08&$g^{D_1}_{D_{s1}D_1K^*}(Q^2)$&3.66&20.89\\
        \end{tabular}
    \end{ruledtabular}
\end{table}
\begin{table}[th]
    \caption{The coupling constant of the vertices ${D^*_sD^*K^*}$ and ${D_{s1}D_1K^*}$,  in $SU_{f}(3)$ symmetry. }\label{T36}
    \begin{ruledtabular}
        \begin{tabular}{ccccccccc}
            $g$&$\mbox{off-shell charmed }$&$\mbox{ off-shell $K^*$}$&$g$&$\mbox{ off-shell charmed}$&$\mbox{ off-shell $K^*$}$\\
            \hline
            $g_{D^*_sD^*K^*}$&$4.88\pm0.64$&$5.03\pm0.65$&$g_{D_{s1}D_1K^*}$&$4.85\pm0.63$&$4.75\pm0.62$\\
        \end{tabular}
    \end{ruledtabular}
\end{table}
Estimated coupling constants of the vertices  ${D^*_sD^*K^*}$, and ${D_{s1}D_1K^*}$, considering the $SU_{f}(3)$ symmetry, are summarized in Table \ref{T36}.
The comparison of the coupling constants $g_{D^*_sD^*K^*}$  with $g_{D^*D^*\rho}$, considering other methods described in references \cite{FSNavarra1,MEBracco}, are given in Table \ref{T37}. It is found that the results of the calculated parameters are in reasonable agreement with that of the references \cite{FSNavarra1,MEBracco} and a factor of two order of magnitude larger in comparison with the references \cite{Lin11,Song11}.
\begin{table}[th]
    \caption{Values of the strong coupling constant reporting different Reference of the  coupling constant $g_{D^*D^*\rho}$ \cite{FSNavarra1,MEBracco,Lin11,Song11}.} \label{T37}
    \begin{ruledtabular}
        \begin{tabular}{cccc}
            $g$ & Ours &Reference \cite{FSNavarra1,MEBracco}& Reference \cite{Lin11,Song11}
            \\ \hline
            $g_{D^*_sD^*K^*}$ &$4.95\pm0.64$&$6.60\pm0.30$ & $2.52$\\
        \end{tabular}
    \end{ruledtabular}
\end{table}

\section{Conclusion}

strong form factors and coupling constants of  $D^*_{s} D^{*} K^*$
and $D_{s1} D_{1} K^*$ vertices were calculated in the frame work of
3-point sum rules of Quantum chromodynamics with and without
consideration of the $SU_{f}(3)$ symmetry. Considering
non-perturbative contributions of the correlation functions, the
quark-quark, quark-gluon, and gluon-gluon condensate corrections
were estimated as the most effective terms. It was found from the
numerical results that the obtained coupling constants are in good
agreement with the other prediction methods described in references
\cite{FSNavarra1,MEBracco}.

\newpage
\section*{Conflict of interest}
The authors of the manuscript declare that there is no conflict of interest regarding publication of this article.

\clearpage
\appendix
\begin{center}
{\Large \textbf{Appendix--A}}
\end{center}
\setcounter{equation}{0} \renewcommand{\theequation}

In this appendix, the explicit expressions of spectral densities are given as:
\begin{eqnarray*}
    \rho^{D^*}_{ D^*_s D^* K^*}=3I_0[3m_s^2-2m_cm_s-s-\Delta+4 A+(C_1-C_2)(2u+2m_cm_s-2s)
    -8(E_1- E_2)]\nonumber \\
    \rho^{K^*}_{ D^*_s D^* K^* }=3I_0[3m_c^2-2m_cm_s-s-\Delta'+4 A'+2(C'_1-C'_2)(u+2m_cm_s-2s)
-8(E'_1- E'_2)]\nonumber \\
\rho^{D_1}_{ D_{s1} D_1 K^*}=3I_0[3m_s^2+2m_cm_s-s-\Delta+4
A+(C_1-C_2)(2u-2m_cm_s-2s)
-8(E_1- E_2)]\nonumber \\
\rho^{K^*}_{ D_{s1} D_1 K^* }=3I_0[3m_c^2+2m_cm_s-s-\Delta'+4
A'+2(C'_1-C'_2)(u-2m_cm_s-2s)
-8(E'_1- E'_2)]
\end{eqnarray*}
Where coefficients in the spectral densities are given as:
\begin{eqnarray*}
    I_0(s,s',q^2) &=& \frac{1}{4\lambda^\frac{1}{2}(s,s',q^2)},\nonumber \\
    \lambda(a,b,c) &=& a^2+ b^2+ c^2- 2ac- 2bc- 2ac ,\nonumber \\
    \Delta&=&s'+m_s^2-m_c^2,\nonumber \\
    \Delta'&=&s'+m_c^2-m_s^2,\nonumber \\
    \Delta'' &=& s+m_s^2,\nonumber \\
    u &=& s+s'-q^2,\nonumber \\
    C_1 &=& \frac{1}{\lambda(s,s',q^2)}[2 s' \Delta'' -u \Delta],\nonumber \\
    C_2 &=& \frac{1}{\lambda(s,s',q^2)}[2 s \Delta -u \Delta'' ],\nonumber \\
    A &=& -\frac{1}{2\lambda(s,s',q^2)}[4ss'm_s^2-s\Delta^2-s'\Delta''^2-m_s^2 u^2+u\Delta\Delta''],\nonumber \\
    E_1 &=& \frac{1}{2\lambda^2(s,s',q^2)}[ 8 ss'^2 m_s^2\Delta''
    -2 s'm_s^2 u^2 \Delta''-4  s s' m_s^2 u \Delta  + m_s^2 u^3 \Delta -2 s'^2 \Delta''^3 \nonumber\\
    &+& 3 s' u \Delta \Delta''^2  -2s s' \Delta^2 \Delta''
    -u^2 \Delta^2 \Delta'' +s u \Delta^3],\nonumber\\
    E_2 &=& \frac{1}{2\lambda^2(s,s',q^2)}[ 8 s^2 s' m_s^2 \Delta-2 s
    m_s^2  u^2 \Delta''- 4  s s' m_s^2 u \Delta'' +m_s^2 u ^3\Delta''-2 s^2 \Delta ^3  \nonumber\\
    &+& 3 s u  \Delta^2 \Delta''-2 s s' \Delta \Delta''^2 -u^2 \Delta
    \Delta''^2 +  s' u \Delta''^3  ],
\end{eqnarray*}
also $A'=A_{|_{m_c\leftrightarrow m_s}},
C'_1={C_1}_{|_{m_c\leftrightarrow m_s}},
C'_2={C_2}_{|_{m_c\leftrightarrow m_s}},
E'_1={E_1}_{|_{m_c\leftrightarrow m_s}}$, and
$E'_2={E_2}_{|_{m_c\leftrightarrow m_s}}$.

\clearpage
\appendix
\begin{center}
    {\Large \textbf{Appendix--B}}
\end{center}
\setcounter{equation}{0} \renewcommand{\theequation}

In this appendix, the explicit expressions of the coefficients of
the quark and gluon condensate contributions of the strong form
factors in the Borel transform scheme for all the vertices are
presented.
\begin{eqnarray*}
    C_{D^*_{s}D^*K^*}^{D^*}&=&\Bigg(-6\,{\frac
        {m_{{s}}{m_{{c}}}^{2}}{{M_{{2}}}^{2}}}-2\,{\frac {{m_{{0}}}
            ^{2}m_{{c}}}{{M_{{2}}}^{2}}}+6\,{\frac
        {m_{{c}}{m_{{s}}}^{2}}{{M_{{2}} }^{2}}}+3\,{\frac
        {m_{{s}}{q}^{2}}{{M_{{2}}}^{2}}}+3\,{\frac {m_{{c}}{
                q}^{2}{m_{{s}}}^{2}}{{M_{{1}}}^{2}{M_{{2}}}^{2}}}-{\frac
        {{m_{{0}}}^{2
            }m_{{c}}{q}^{2}}{{M_{{1}}}^{2}{M_{{2}}}^{2}}}-3\,{\frac
        {{m_{{c}}}^{3}
            {m_{{s}}}^{2}}{{M_{{1}}}^{2}{M_{{2}}}^{2}}}\\&+&{\frac
        {{m_{{0}}}^{2}{m_{{ c}}}^{3}}{{M_{{1}}}^{2}{M_{{2}}}^{2}}}-3\,{\frac
        {{m_{{c}}}^{3}{m_{{s} }}^{2}}{{M_{{2}}}^{4}}}+\frac{3}{2}\,{\frac
        {{m_{{0}}}^{2}{m_{{c}}}^{3}}{{M_{{ 2}}}^{4}}}+3\,m_{{s}}\Bigg)\times
    e^{-\frac{m_c^2}{M_2^2}},
\end{eqnarray*}
\begin{eqnarray*}
    C_{D_{s1}D_1K^*}^{D_1}&=& \Bigg(2\,{\frac
        {{m_{{0}}}^{2}m_{{c}}}{{M_{{2}}}^{2}}}-6\,{\frac {m_{{s}}{m_
                {{c}}}^{2}}{{M_{{2}}}^{2}}}+3\,{\frac
        {m_{{s}}{q}^{2}}{{M_{{2}}}^{2}}} -6\,{\frac
        {m_{{c}}{m_{{s}}}^{2}}{{M_{{2}}}^{2}}}-{\frac {{m_{{0}}}^{2
            }{m_{{c}}}^{3}}{{M_{{1}}}^{2}{M_{{2}}}^{2}}}-3\,{\frac
        {m_{{c}}{q}^{2} {m_{{s}}}^{2}}{{M_{{1}}}^{2}{M_{{2}}}^{2}}}+{\frac
        {{m_{{0}}}^{2}m_{{c
            }}{q}^{2}}{{M_{{1}}}^{2}{M_{{2}}}^{2}}}\\&+&3\,{\frac
        {{m_{{c}}}^{3}{m_{{s }}}^{2}}{{M_{{1}}}^{2}{M_{{2}}}^{2}}}+3\,{\frac
        {{m_{{c}}}^{3}{m_{{s}} }^{2}}{{M_{{2}}}^{4}}}-\frac{3}{2}\,{\frac
        {{m_{{0}}}^{2}{m_{{c}}}^{3}}{{M_{{2 }}}^{4}}}+3\,m_{{s}}\Bigg)\times
    e^{-\frac{m_c^2}{M_2^2}},
\end{eqnarray*}

\begin{eqnarray*}
    C_{D^*_{s}D^*K^*}^{K^*}&=&\hat{I}_{0}(3,2,2)m_{c}^{6}-\hat{I}_{0}(3,2,2)m_{c}^{5}m_{s}+\hat{I}_{0}(3,2,2)m_{c}^{3}m_{s}^{3}+\hat{I}_1^{[1,0]}(3,2,2)m_{c}^{4}
    \\
    &&+3\hat{I}_{0}(4,1,1)m_{c}^{4}-\hat{I}_2^{[1,0]}(3,2,2)m_{c}^{4}+2\hat{I}_{6}(3,2,2)m_{c}^{4}+2\hat{I}_{2}(2,1,3)m_{c}^{3}m_{s}
    \\
    &&+2\hat{I}_{0}(2,1,3)m_{c}^{3}m_{s}-2\hat{I}_{1}(2,2,2)m_{c}^{3}m_{s}-2\hat{I}_{1}(2,1,3)m_{c}^{3}m_{s}-\hat{I}_{1}(3,2,1)m_{c}^{3}m_{s}
    \\
    &&+\hat{I}_{2}(3,2,1)m_{c}^{3}m_{s}-3\hat{I}_{0}(4,1,1)m_{c}^{3}m_{s}+2\hat{I}_{2}(2,2,2)m_{c}^{3}m_{s}+\hat{I}_{0}(2,2,2)m_{c}^{2}m_{s}^{2}
    \\
    &&-2\hat{I}_{6}(3,2,2)m_{c}^{2}m_{s}^{2}+\hat{I}_0^{[0,1]}(3,2,2)m_{c}^{2}m_{s}^{2}+\hat{I}_2^{[1,0]}(3,2,2)m_{c}^{2}m_{s}^{2}-\hat{I}_1^{[1,0]}(3,2,2)m_{c}^{2}m_{s}^{2}
    \\
    &&+2\hat{I}_{2}(3,1,2)m_{c}m_{s}^{3}-\hat{I}_2^{[1,0]}(3,2,2)m_{c}m_{s}^{3}-2\hat{I}_{1}(3,1,2)m_{c}m_{s}^{3}+6\hat{I}_{0}(1,1,4)m_{c}m_{s}^{3}
    \\
    &&+\hat{I}_1^{[1,0]}(3,2,2)m_{c}m_{s}^{3}+\hat{I}_{0}(3,1,2)m_{s}^{4}+2\hat{I}_6^{[1,0]}(3,2,2)m_{c}^{2}+\hat{I}_{0}(2,1,2)m_{c}^{2}
    \\
    &&+8\hat{I}_{8}(3,2,1)m_{c}^{2}+4\hat{I}_8^{[0,1]}(3,2,2)m_{c}^{2}+2\hat{I}_6^{[0,1]}(3,2,2)m_{c}^{2}-4\hat{I}_{6}(3,2,1)m_{c}^{2}
    \\
    &&-4\hat{I}_7^{[0,1]}(3,2,2)m_{c}^{2}+3\hat{I}_1^{[1,0]}(4,1,1)m_{c}^{2}+6\hat{I}_{6}(4,1,1)m_{c}^{2}+6\hat{I}_{6}(3,1,2)m_{c}^{2}
    \\
    &&-8\hat{I}_{7}(3,2,1)m_{c}^{2}-12\hat{I}_{7}(4,1,1)m_{c}^{2}+2\hat{I}_{0}(2,2,1)m_{c}^{2}+12\hat{I}_{8}(4,1,1)m_{c}^{2}
    \\
    &&-3\hat{I}_2^{[1,0]}(4,1,1)m_{c}^{2}+8\hat{I}_{6}(2,1,3)m_{c}m_{s}+4\hat{I}_{6}(2,2,2)m_{c}m_{s}-\hat{I}_0^{[1,1]}(3,2,2)m_{c}m_{s}
    \\
    &&-3\hat{I}_{1}(1,3,1)m_{c}m_{s}-2\hat{I}_1^{[0,1]}(3,1,2)m_{c}m_{s}+3\hat{I}_1^{[1,0]}(3,2,1)m_{c}m_{s}-2\hat{I}_{6}(3,2,1)m_{c}m_{s}
    \\
    &&+3\hat{I}_{2}(1,3,1)m_{c}m_{s}+2\hat{I}_2^{[0,1]}(3,1,2)m_{c}m_{s}-\hat{I}_{0}(2,1,2)m_{c}m_{s}-2\hat{I}_{6}(3,1,2)m_{c}m_{s}
    \\
    &&-3\hat{I}_2^{[1,0]}(3,2,1)m_{c}m_{s}-2\hat{I}_6^{[1,0]}(3,2,2)m_{s}^{2}-12\hat{I}_{6}(1,1,4)m_{s}^{2}-\hat{I}_{0}(2,1,2)m_{s}^{2}
    \\
    &&-3\hat{I}_{0}(1,1,3)m_{s}^{2}-2\hat{I}_0^{[0,1]}(3,1,2)m_{s}^{2}+3\hat{I}_0^{[1,0]}(1,1,4)m_{s}^{2}+\hat{I}_0^{[1,1]}(3,2,2)m_{s}^{2}
    \\
    &&+6\hat{I}_2^{[1,0]}(1,1,4)m_{s}^{2}+\hat{I}_{0}(3,1,1)m_{s}^{2}+24\hat{I}_{7}(1,1,4)m_{s}^{2}-24\hat{I}_{8}(1,1,4)m_{s}^{2}
    \\
    &&+4\hat{I}_{6}(2,2,2)m_{s}^{2}-6\hat{I}_1^{[1,0]}(1,1,4)m_{s}^{2}-2\hat{I}_0^{[1,0]}(2,2,2)m_{s}^{2}+3\hat{I}_0^{[0,1]}(1,1,4)m_{s}^{2}
    \\
    &&+2\hat{I}_{6}(3,1,2)m_{s}^{2}+3\hat{I}_1^{[0,1]}(2,1,2)-2\hat{I}_{2}(2,1,1)+3\hat{I}_{6}(1,3,1) \\
    &&-3\hat{I}_2^{[0,1]}(2,1,2)-4S_{1},1(3,2,2)+4\hat{I}_7^{[1,0]}(3,2,1)-8\hat{I}_{8}(2,2,1) \\
    &&-2\hat{I}_6^{[0,1]}(2,2,2)-2\hat{I}_6^{[0,1]}(3,1,2)+8\hat{I}_{7}(2,2,1)-4\hat{I}_8^{[1,0]}(3,2,1) \\
    &&-4\hat{I}_6^{[1,0]}(3,2,1)+2\hat{I}_6^{[1,1]}(3,2,2)-12\hat{I}_{8}(3,1,1)+3\hat{I}_1^{[1,0]}(3,1,2) \\
    &&+2\hat{I}_{1}(2,1,1)-2\hat{I}_2^{[1,0]}(1,2,2)+2\hat{I}_1^{[1,0]}(1,2,2)+4\hat{I}_7^{[0,1]}(2,2,2) \\
    &&-4\hat{I}_8^{[0,1]}(2,2,2)+4\hat{I}_{6}(2,2,1)-\hat{I}_0^{[0,1]}(3,1,1)-3\hat{I}_2^{[1,1]}(3,1,2) \\
    &&+\hat{I}_2^{[2,0]}(2,2,2)-\hat{I}_1^{[2,0]}(2,2,2)+12\hat{I}_{7}(3,1,1)-8\hat{I}_{6}(2,1,2) \\
    &&+4\hat{I}_8^{[1,1]}(3,2,2)+\hat{I}_0^{[2,0]}(3,2,1)-2\hat{I}_0^{[1,0]}(1,2,2)-\hat{I}_0^{[0,1]}(2,1,2) \\
    &&-3\hat{I}_{0}(2,1,1)-4\hat{I}_{6}(3,1,1)-2\hat{I}_6^{[1,0]}(2,2,2)+4\hat{I}_7^{[1,0]}(2,2,2) \\
    &&-4\hat{I}_8^{[1,0]}(2,2,2)+2\hat{I}_{0}(1,1,2)-\hat{I}_0^{[1,0]}(3,1,1)-6\hat{I}_{7}(1,3,1)\\
    &&+6\hat{I}_{8}(1,3,1),
\end{eqnarray*}
\begin{eqnarray*}
    C_{D_{s1}D_1K^*}^{K^*}&=&\hat{I}_{0}(3,2,2)m_{c}^{6}+\hat{I}_{2}(3,2,2)m_{c}^{3}m_{s}^{3}-\hat{I}_{1}(3,2,2)m_{c}^{3}m_{s}^{3}-\hat{I}_{0}(3,2,2)m_{c}^{3}m_{s}^{3}
    \\
    &&-4\hat{I}_{7}(3,2,2)m_{c}^{4}+3\hat{I}_{0}(2,2,2)m_{c}^{4}+2\hat{I}_{6}(3,2,2)m_{c}^{4}+4\hat{I}_{8}(3,2,2)m_{c}^{4}
    \\
    &&+3\hat{I}_{0}(4,1,1)m_{c}^{4}-2\hat{I}_{2}(2,2,2)m_{c}^{3}m_{s}+2\hat{I}_{1}(2,2,2)m_{c}^{3}m_{s}-2\hat{I}_{2}(2,1,3)m_{c}^{3}m_{s}
    \\
    &&+2\hat{I}_{1}(2,1,3)m_{c}^{3}m_{s}+\hat{I}_{1}(3,2,1)m_{c}^{3}m_{s}+\hat{I}_0^{[0,1]}(3,2,2)m_{c}^{3}m_{s}-\hat{I}_{2}(3,2,1)m_{c}^{3}m_{s}
    \\
    &&-\hat{I}_{0}(3,1,2)m_{c}^{3}m_{s}-2\hat{I}_{6}(3,2,2)m_{c}^{2}m_{s}^{2}-\hat{I}_1^{[1,0]}(3,2,2)m_{c}^{2}m_{s}^{2}-6\hat{I}_{0}(1,1,4)m_{c}^{2}m_{s}^{2}
    \\
    &&+\hat{I}_2^{[1,0]}(3,2,2)m_{c}^{2}m_{s}^{2}+6\hat{I}_{2}(1,1,4)m_{c}m_{s}^{3}-6\hat{I}_{1}(1,1,4)m_{c}m_{s}^{3}+\hat{I}_{0}(3,1,2)m_{s}^{4}
    \\
    &&+\hat{I}_{1}(2,1,2)m_{c}^{2}-\hat{I}_{1}(3,1,1)m_{c}^{2}+2\hat{I}_{0}(2,2,1)m_{c}^{2}+\hat{I}_1^{[2,0]}(3,2,2)m_{c}^{2}
    \\
    &&+8\hat{I}_{8}(3,2,1)m_{c}^{2}+4\hat{I}_8^{[0,1]}(3,2,2)m_{c}^{2}+2\hat{I}_6^{[0,1]}(3,2,2)m_{c}^{2}-12\hat{I}_{7}(4,1,1)m_{c}^{2}
    \\
    &&-4\hat{I}_7^{[0,1]}(3,2,2)m_{c}^{2}+4\hat{I}_8^{[1,0]}(3,2,2)m_{c}^{2}+\hat{I}_{2}(3,1,1)m_{c}^{2}+2\hat{I}_8^{[1,0]}(3,2,2)m_{c}^{2}
    \\
    &&-\hat{I}_2^{[2,0]}(3,2,2)m_{c}^{2}-4\hat{I}_{6}(3,2,1)m_{c}^{2}-8\hat{I}_{7}(3,2,1)m_{c}^{2}-\hat{I}_0^{[0,1]}(2,2,2)m_{c}^{2}
    \\
    &&+\hat{I}_{0}(2,1,2)m_{c}^{2}+6\hat{I}_{6}(4,1,1)m_{c}^{2}-\hat{I}_{2}(2,1,2)m_{c}^{2}-4\hat{I}_7^{[1,0]}(3,2,2)m_{c}^{2}
    \\
    &&+6\hat{I}_{6}(3,1,2)m_{c}^{2}+12\hat{I}_{8}(4,1,1)m_{c}^{2}-4\hat{I}_{1}(1,1,3)m_{c}m_{s}+4\hat{I}_{2}(1,1,3)m_{c}m_{s}
    \\
    &&-4\hat{I}_{6}(2,2,2)m_{c}m_{s}-3\hat{I}_{2}(1,3,1)m_{c}m_{s}+2\hat{I}_{0}(1,2,2)m_{c}m_{s}+2\hat{I}_{6}(3,1,2)m_{c}m_{s}
    \\
    &&+\hat{I}_{0}(2,1,2)m_{c}m_{s}-2\hat{I}_2^{[0,1]}(3,1,2)m_{c}m_{s}-8\hat{I}_{6}(2,1,3)m_{c}m_{s}+2\hat{I}_0^{[1,0]}(2,1,3)m_{c}m_{s}
    \\
    &&+2\hat{I}_{6}(3,2,1)m_{c}m_{s}+\hat{I}_{2}(2,1,2)m_{c}m_{s}+2\hat{I}_1^{[0,1]}(3,1,2)m_{c}m_{s}-\hat{I}_{1}(2,1,2)m_{c}m_{s}
    \\
    &&+3\hat{I}_{1}(1,3,1)m_{c}m_{s}+3\hat{I}_{0}(2,2,1)m_{c}m_{s}-3\hat{I}_{0}(1,1,3)m_{s}^{2}+4\hat{I}_{6}(2,2,2)m_{s}^{2}
    \\
    &&-6\hat{I}_1^{[1,0]}(1,1,4)m_{s}^{2}+3\hat{I}_0^{[0,1]}(1,1,4)m_{s}^{2}-12\hat{I}_{6}(1,1,4)m_{s}^{2}-2\hat{I}_0^{[0,1]}(3,1,2)m_{s}^{2}
    \\
    &&-\hat{I}_{0}(2,1,2)m_{s}^{2}+6\hat{I}_2^{[1,0]}(1,1,4)m_{s}^{2}+2\hat{I}_{6}(3,1,2)m_{s}^{2}-2\hat{I}_6^{[1,0]}(3,2,2)m_{s}^{2}
    \\
    &&-2\hat{I}_2^{[1,0]}(1,2,2)+\hat{I}_2^{[1,0]}(3,1,1)-4\hat{I}_6^{[1,0]}(3,2,1)-\hat{I}_0^{[1,0]}(3,1,1) \\
    &&-8\hat{I}_{8}(2,2,1)-\hat{I}_0^{[0,1]}(2,1,2)-12\hat{I}_{8}(3,1,1)-2\hat{I}_6^{[1,0]}(2,2,2)+12\hat{I}_{7}(3,1,1)
    \\
    &&-8\hat{I}_{8}(2,1,2)-8\hat{I}_{6}(2,1,2)-\hat{I}_1^{[1,0]}(3,1,1)+8\hat{I}_{7}(2,1,2)+4\hat{I}_7^{[1,0]}(3,2,1)
    \\
    &&-4\hat{I}_8^{[1,0]}(3,2,1)+4\hat{I}_7^{[0,1]}(3,1,2)-4\hat{I}_8^{[0,1]}(3,1,2)+2\hat{I}_1^{[0,1]}(1,2,2) \\
    &&+2\hat{I}_{1}(2,1,1)-4\hat{I}_7^{[1,1]}(3,2,2)+4\hat{I}_8^{[1,1]}(3,2,2)+3\hat{I}_{6}(1,3,1)-2\hat{I}_6^{[0,1]}(3,1,2)
    \\
    &&-2\hat{I}_6^{[0,1]}(2,2,2)-4\hat{I}_{6}(3,1,1)+\hat{I}_0^{[0,2]}(3,1,2)-\hat{I}_0^{[0,1]}(3,1,1) \\
    &&-3\hat{I}_{0}(2,1,1)+4\hat{I}_{6}(2,2,1)+2\hat{I}_6^{[1,1]}(3,2,2)-2\hat{I}_{2}(2,1,1) \\
    &&+8\hat{I}_{7}(2,2,1)-2\hat{I}_0^{[0,1]}(2,2,1),
\end{eqnarray*}

where
\begin{eqnarray*}
    \hat{I}_{\mu}^{[\alpha,\beta]} (a,b,c)&=&
    [M_1^2]^{\alpha}[M_2^2]^{\beta}\frac{d^\alpha}{d(M_1^2)^{\alpha}}
    \frac{d^\beta}{d(M_2^2)^{\beta}}[M_1^2]^{\alpha}[M_2^2]^{\beta}\hat
    I_{\mu} (a,b,c), \nonumber \\ \hat{I}_k(a,b,c) \!\!\! &=& \!\!\! i
    \frac{(-1)^{a+b+c+1}}{16 \pi^2\,\Gamma(a) \Gamma(b) \Gamma(c)}
    (M_1^2)^{1-a-b+k} (M_2^2)^{4-a-c-k} \, {U}_0(a+b+c-5,1-c-b),
    \nonumber \\ \hat{I}_m(a,b,c) \!\!\! &=& \!\!\! i
    \frac{(-1)^{a+b+c+1}}{16 \pi^2\,\Gamma(a) \Gamma(b) \Gamma(c)}
    (M_1^2)^{-a-b-1+m} (M_2^2)^{7-a-c-m} \, {U}_0(a+b+c-5,1-c-b),
    \nonumber\\ \hat{I}_6(a,b,c) \!\!\! &=& \!\!\! i
    \frac{(-1)^{a+b+c+1}}{32 \pi^2\,\Gamma(a) \Gamma(b) \Gamma(c)}
    (M_1^2)^{3-a-b} (M_2^2)^{3-a-c} \, {U}_0(a+b+c-6,2-c-b), \nonumber\\
    \hat{I}_n(a,b,c) \!\!\! &=& \!\!\! i \frac{(-1)^{a+b+c}}{32
        \pi^2\,\Gamma(a) \Gamma(b) \Gamma(c)} (M_1^2)^{-4-a-b+n}
    (M_2^2)^{11-a-c-n} \, {U}_0(a+b+c-7,2-c-b),
\end{eqnarray*}
where $k=1, 2$, $m=3, 4, 5$ and $n=7, 8$. We can define the function
$U_0(a,b)$ as:
\begin{eqnarray*}
    U_0 (a, b) = \int_0^{\infty} dy (y + M_1^2 + M_2^2)^ay^b \exp
    [-\frac{B_{-1}}{y} - B_0 - B_1y ],
\end{eqnarray*}
where
\begin{eqnarray*}
    B_{-1} &=& \frac{1}{M_2^2M_1^2}(m_s^2(M_1^2 +M_2^2)^2
    -M_2^2M_1^2Q^2), \nonumber\\
    B_{0} &=& \frac{1}{M_1^2M_2^2}(m_s^2 + m_c^2)(M_1^2+M_2^2)
    ,\\
    B_{1} &=& \frac{m_c^2}{M_1^2M_2^2}. \nonumber
\end{eqnarray*}

%%%%%%%%%%%%%%%%%%%%%%%%%%%%%%%%%%%%%%%%%%%%%%%%%%%%%%%%%%%%%%%%%%%%%%%%%%%%%%%%%%%%%%%%%%%%%%%%%%%%%%%%%%%%%%


\begin{thebibliography}{II}

\bibitem{exotic}
M. L. Du, W. Chen, X. L. Chen and S. L. Zhu, \textit{Phys. Rev. D} \textbf{87}, 014003 (2013).

\bibitem{MEBracco}
M. E. Bracco, M. Chiapparini, F. S. Navarra, and M. Nielsen, \textit{Phys.
Lett. B} \textbf{659}, 559 (2008).

\bibitem{FSNavarra1}
F. S. Navarra, M. Nielsen, M. E. Bracco, M. Chiapparini, C. L.
Schat, \textit{Phys. Lett. B} \textbf{489}, 319 (2000).

\bibitem{MNielsen}
F. S. Navarra, M. Nielsen, and M. E. Bracco, \textit{Phys. Rev. D}
\textbf{65}, 037502 (2002).

\bibitem{MChiapparini}
M. E. Bracco, M. Chiapparini, A. Lozea, F. S. Navarra, and M.
Nielsen, \textit{Phys. Lett. B} \textbf{521}, 1 (2001).

\bibitem{Rodrigues3}
B. O. Rodrigues, M. E. Bracco, M. Nielsen, and F. S. Navarra, \textit{Nucl.
Phys. A} \textbf{852}, 127 (2011).

\bibitem{RDMatheus}
R. D. Matheus, F. S. Navarra, M. Nielsen, and R. R. da Silva, \textit{Phys.
Lett. B} \textbf{541}, 265 (2002).

\bibitem{RRdaSilva}
R. R. Da Silva, R. D. Matheus, F. S. Navarra, and M. Nielsen,  \textit{Braz.
J. Phys.} \textbf{34}, 236 (2004).

\bibitem{SLWang}
Z. G. Wang, and S. L. Wan, \textit{Phys. Rev. D} \textbf{74}, 014017 (2006).

\bibitem{ZGWang}
Z. G. Wang, \textit{Nucl. Phys. A} \textbf{796}, 61 (2007).

\bibitem{FCarvalho}
F. Carvalho, F. O. Duraes, F. S. Navarra, and M. Nielsen,  \textit{Phys.
Rev. C} \textbf{72}, 024902 (2005).

\bibitem{ALozea}
M. E. Bracco, A. J. Cerqueira, M. Chiapparini, A. Lozea, and M.
Nielsen, \textit{Phys. Lett. B} \textbf{641}, 286 (2006).

\bibitem{LBHolanda}
L. B. Holanda, R. S. Marques de Carvalho, and A. Mihara,  \textit{Phys.
Lett. B} \textbf{644}, 232 (2007).

\bibitem{Wang122}
Z. G. Wang, and S. L. Wan, \textit{Phys. Rev. D} \textbf{74}, 014017
(2006).

\bibitem{Wang3434}
Z. G. Wang, \textit{J. Phys. G} \textbf{34}, 753 (2007).

\bibitem{Ghahramany}
N. Ghahramany, R. Khosravi, and M. Janbazi,  \textit{Int. J. Mod. Phys. A} \textbf{27}, 1250022 (2012).

\bibitem{KJ}
R. Khosravi, and M. Janbazi, \textit{Phys. Rev. D} \textbf{87}, 016003 (2013).

\bibitem{KJ2}
R. Khosravi1, and M. Janbazi, \textit{Phys. Rev. D} \textbf{89}, 016001 (2014).

\bibitem{Janbazi}
M. Janbazi, N. Ghahramany, and E. Pourjafarabadi, \textit{Eur. Phys. J. C}
\textbf{74}, 2718 (2014).

\bibitem{BraChia}
M. E. Bracco, M. Chiapparini, F. S. Navarra and M. Nielsen,
\textit{Prog. Part. Nucl. Phys.} \textbf{67}, 1019 (2011),  arXiv:1104.2864 [hep-ph].

\bibitem{Colangelo}
P. Colangelo, and A. Khodjamirian, arXiv:0010175 [hep-ph].

\bibitem{PDG2012}
J. Beringer \textit{et al.}, \textit{Phys. Rev. D} \textbf{86}, 010001 (2012).

\bibitem{Gelhausen}
P. Gelhausen, A. Khodjamirian, A. A. Pivovarov, and D. Rosenthal,
\textit{et al.}, \textit{Phys. Rev. D} \textbf{88}, 014015 (2013).

\bibitem{Thoma}
C. E. Thomas, \textit{Phys. Rev. D} \textbf{73}, 054016 (2006).

\bibitem{Bazavov}
A. Bazavov \textit{et al.}, \textit{Phys. Rev. D} \textbf{85}, 114506 (2012).

\bibitem{Bracco 1111}
M. E. Bracco, \textit{et al.} \textit{Prog. Part. Nucl. Phys.} \textbf{67} 1019 (2012).


\bibitem{Rodrigues 111}
B. O. Rodrigues, B. Osório, M. E. Bracco, and A. Cerqueira Jr, \textit{Nuclear Physics A} \textbf{957} 109 (2017).

\bibitem{Lin11}
Z. Lin, and C. M. Ko, \textit{Phys. Rev. C} \textbf{62}, 034903
(2000); Z. Lin, C. M. Ko and B. Zhang, \textit{Phys. Rev. C}
\textbf{61}, 024904 (2000).

\bibitem{Song11}
Y. Oh, T. Song and S. H. Lee, \textit{Phys. Rev. C} \textbf{63},
034901 (2001).



\end{thebibliography}
\end{document}